\documentstyle[nato]{crckapb}

\newcommand{\nfig}[6] %
{ \begin{figure}[#6]
  \footnotesize
  \begin{center}
   \unitlength1cm
     \begin{picture}(#3,#4)
     \end{picture}
  \refstepcounter{figure}\label{#1}
  \vskip 14pt
  \begin{minipage}{#5cm}

  {\bf Fig.~\thefigure .} ~~#2

  \end{minipage}
  \end{center}
  \normalsize
  \end{figure} }

\newcommand{\sfig}[7] %
{\begin{figure}[#7]
 \unitlength1cm
 \small
  \begin{minipage}[t]{#3cm}
     \begin{picture}(#4,#5)
     \end{picture}
  \hfill
     \begin{minipage}[b]{#6cm}
     \refstepcounter{figure}\label{#1}

     {\bf Fig.~\thefigure .}~~#2

     \end{minipage}
  \end{minipage}
  \normalsize
 \end{figure} }
\begin{opening}
\title{LOCAL STRUCTURE STUDIES OF THE\protect\\ 
UNDERDOPED-OVERDOPED TRANSITION IN 
$\rm\bf YB{\lowercase{a}}_2C{\lowercase{u}}_3O_{\lowercase{x}}$}

\subtitle{Measurement of the yttrium x-ray absorption-fine-structure}

\author{J. R\"OHLER$^1$}
\author{P. W. LOEFFEN}
\author{S. M\"ULLENDER}
\institute{European Synchrotron Radiation Facility,\\
           B.P. 220, F-38043, France}
\author{K. CONDER}
\author{E. KALDIS}
\institute{Laboratorium f\"ur Festk\"orperphysik ETH,\\
          CH-8093 Z\"urich, Switzerland}
\end{opening}

\runningtitle{Y-EXAFS}

\begin{document}
\footnotetext[1]{Corresponding author. 
Universit{\"a}t zu K{\"o}ln, II. Physikalisches
Institut, Z{\"u}lpicherstr. 77, D-50937 K{\"o}ln, Germany. 
Email:abb12@RS1.RRZ.Uni-Koeln.de}

\begin{abstract}
We have measured the extended x-ray absorption-fine-structure (EXAFS)
at the Y-$K$ edge of $\rm YBa_2Cu_3O_x$ for $x$\/=6.801, 6.947, 6.968,
6.984 at $T$\/=20-300 K. The Y--Cu2 pairs vibrate harmonically but
freeze out in the superconducting phase. The Y-O2,3 pairs exhibit
strong anharmonicities with a singularity at $T_c$. With increasing
oxygen concentration the Cu2 layer shifts along $c$ towards the Ba
layer. Optimum doping is a notable point in the phase diagram, also
concerning the structural degrees of freedom. Here the O2,3--Cu2
spacing is largest, the relative displacements of O2,3 and Cu2 layers
along $c$ invert upon doping, and the Cu2 position along $c$ is
independent on temperature in the superconducting phase.
\end{abstract}

\section{\label{secIntro}Introduction}

Upon doping of the superconducting cuprates, one distinguishes an
under- and overdoped regime. Optimum doping defines the
unique point in the phase diagram where the critical
temperature, $T_c$, is at a maximum. The superconducting phase of
$\rm YBa_2Cu_3O_x$ exhibits a broad but well defined maximum of
$T_c$ in the oxygen concentration range $6.8<x<7.0$
\cite{ConJil,ClaLoe}. From $T_c$ $vs.$ $x$ of homogeneously
oxidized powder samples \cite{ConJil} we read the
maximum $T_c=92.5$ K is at $x_{opt}=6.92\pm 0.01$.

Optimum doped $\rm YBa_2Cu_3O_{6.92}$ contrasts with its under- and
overdoped neighbours not only by the relatively highest $T_c$, but
also by the narrowest width in the superconducting transition (in the
electrical resistance, magnetization, specific heat), and the largest
superconducting volume fraction. Moreover, at $x_{opt}$ the anisotropy
ratio of all electrical and magnetic properties is largest and the
normal $a$-$b$ resistivity most linear in temperature. It is generally
believed that the maximum of $T_c$ is related to a critical
concentration of charge carriers in the $\rm CuO_2$ planes,
$n_c(x)\simeq 0.2-0.25$ holes/Cu, which may be related to an optimum
superconducting condensate density $\rho_s(0)=n_s(0)/m_{ab}^*$ at zero
temperature. ($m^*$ is the effective mass of the superconducting
carriers in $a,b$ direction.) At optimum doping $\partial T_c/\partial
n\simeq 0$. It is important to note that the oxygen isotope shift is
smallest at optimum doping \cite{ZecMue,ZecCon}. Measurements of the
electrical resistivity under high pressure on $\rm YBa_2Cu_3O_x$ has
also established a vanishing pressure dependence of $T_c$,
$dT_c/dp\simeq 0.04$ K/kbar \cite{AlmSch}. From the currently
available experimental data, $\rm YBa_2Cu_3O_{x}$ turns out to be the
family of cuprate superconductors where optimum doping is most
precisely characterized. 

The other families of cuprate superconductors 
have been well defined in their under- and overdoped 
regimes, but as yet optimum doping could not be
determined with the same accuracy as in $\rm YBa_2Cu_3O_x$.  
Carrier concentration is a key variable
influencing $T_c$, but in the various compound families, control
of the carrier concentration in the $\rm CuO_2$ planes is
achieved in a multi parameter space of cation and anion
concentrations. At fixed compositions of the dopants, 
ordering of the inherent defects has been established as 
playing an important role in the
variation of $T_c$. Even in the structurally ``simple'' compound
$\rm La_{2-x}Sr_xCuO_4$ optimum doping is not completely
understood. For instance, $\rm La_{2-x}Sr_xCuO_4$ exhibits the 
maximum $T_c$ (38 K) at $x$\/=0.15, however, application of
pressure may enhance $T_c$ at just this concentration 
up to 49 K. We believe a key for understanding
of the narrow range of stoichiometries which allows for
optimum doping is in the ordering of the inherent defect
structures. The optimum doped superconducting phases have
intermediate carrier concentrations with respect to the
insulating (antiferromagnetic) parent phases and the overdoped
nonsuperconducting metallic phases. It is an important
indication that just these intermediate carrier concentrations
are most favourable for the superconductivity.

An exceptionally detailed picture of the role of defects in a
high-$T_c$ material emerged from the numerous detailed studies of
oxygen vacancies in the chain layer of underdoped $\rm YBa_2Cu_3O_x$,
$6.4<x<6.92$. For a given $x$, the highest $T_c$ is generally found
for samples annealed so that the interstitial oxygen O4 in the chain
layer is well ordered. The oxygen ordering is manifested as dramatic
increases in $T_c$ with time in samples that have been quenched from
high temperature. Alternatively, application of external pressure to
underdoped $\rm YBa_2Cu_3O_x$ may be used to affect the
ordering-disordering of the chain oxygens. Note that 'ordered defects'
are not defects at all; instead they have become part of the regular
lattice, $e.g.$ the so-called 2$a$ ortho-II superstructure,
\cite{GryKub,ZeiHoh}. 

But many defects remain after the chain oxygen becomes well ordered
creating local lattice distortions. For instance, the copper-oxygen
chains are not straight but zigzagged, and the oxygen displacements
associated with this distortion are not ordered in three dimensions.
It is important to note that even the $\rm CuO_2$ planes, generally
believed to be structurally most perfect, are locally distorted by the
atomically defined dopant sites located in the nearest or next nearest
layers. Are there specific configurations of ordered distortions,
which are most favourable for the superconductivity, and which
stabilize the compound at optimum doping? To elucidate this problem,
experimentally homogeneously oxidized samples and precision
measurements of their oxygen concentrations are necessary
prerequisites.

The overdoped regime exhibits further intriguing properties. The jump
in the specific heat at $T_c$, $\Delta C_p/T_c$ saturates for
$x>x_{opt}$. Compared to underdoped $\rm YBa_2Cu_3O_{6.80}$ it is
enhanced by a factor of three \cite{ClaLoe}. At $x>6.95$ new features
show up in the shapes of the specific heat jumps and of the
magnetization. These are interpreted as a double superconducting
transition with $T_c's$ split by 3 K \cite{LorMir}. The splitting
cannot be explained by macroscopic oxygen inhomogeneities in the
grains \cite{ConKel}. Overdoping tends to supress the superconducting
properties, to dramatically reduce the anisotropy ratio of the normal
electric transport properties, and to cause the normal resistivity to
evolve a power law dependence. 

X-ray and neutron diffraction data recorded as a function of oxygen
concentration at 5 K exhibit clear structural anomalies at $x\geq
6.935$ \cite{ConJil}. Further insertion of oxygen was found to invert
the $x$\/-dependency of the $c$\/ lattice parameter. The spacing
between the oxygen and copper layers in the $\rm CuO_2$ plane, $\Delta
z_{O2,3-Cu2}$, drops at $x\simeq 6.925$ by 0.02 $\rm\AA$. Recently the
in-phase Raman mode of the planar oxygens, O2,3, was found to soften
in the overdoped regime \cite{PouMue}.  

In this lecture, we address the problem of local structural
distortions in the $\rm CuO_2$ planes around optimum doping in
$\rm YBa_2Cu_3O_x$, $x$=6.801-6.984, by measurements of the
extended x-ray absorption-fine-structure (EXAFS) beyond the
yttrium $K$\/-edge, both, as a function of oxygen concentration,
$x$\/=6.801-6.984, and of temperature $T$\/=20-300 K. 

\section{\label{SecExperiment}Experimental Details}
\subsection{\label{SubEXAFS}X-RAY ABSORPTION-FINE-STRUCTURE}
The x-ray absorption-fine-structure beyond the x-ray absorption
thresholds of atoms in solids arises from the interference
between freely propagating photoelectrons and photoelectrons
backscattered from the atomic cluster to the absorbing central
atom. X-ray photons are used for generating EXAFS spectra, but it is
interfering photoelectrons which give the oscillating absorption
pattern used for the analysis of the local structure. Since
electrons are strongly scattered by atoms, EXAFS is a very
sensitive structural probe, but exhibits strong contributions
from complicated multiple scattering (MS). However, strong MS signals
are of great advantage for the detection
\sfig{Fig1} 
{Atomic cluster pro\-jected from the unit cell of $\rm
YBa_2\-Cu_3O_x$. Shown are two $\rm CuO_2$ pla\-quettes (Cu: small
black circles, O: big dotted circles) se\-pa\-rated by the Y-layer
(big black circles). On the top we see one Ba layer (hatched
circles) including the apical oxygen of the Cu--O pyramid.
Arrowed lines indicate sche\-matically some of the important Y EXAFS
scattering paths: Y\---O2,3 (``O2,3''), Y\---Cu2 (''Cu2``), and
Y\---Cu2\---Ba (''6.2``), Y\---O2,3\---Ba (''5``), discussed
in the text.}   
{12.5}     
{8.3168}   
{6.3845}   
{4.}       
{b}        
of subtle structural effects and the improvement of the structural
analysis. On the other hand, inclusion of MS into the data analysis
markedly increases the complexity of the structural refinement, far
beyond that of the so-called standard analysis carried out earlier on
the Y $K$-EXAFS of $\rm YBa_2Cu_3O_x$ by many laboratories
\cite{YanXu,BoyCla,ZhaHua,ZhaGal}. We have recently determined the
so-called dimpling angle in the $\rm CuO_2$ planes of $\rm
BiSr_2Ca_2Cu_2O_{8+\delta}$ \cite{RoeCru} from a MS analysis of the Cu
$K$\/-EXAFS using the FEFF code \cite{RehZab}; MS analysis of the
EXAFS from high-$T_c$ cuprates using other codes have been reported in
$e.g$ Ref. \cite{CicBer,GurJor}.

Inspection of our experimental Fourier transform
spectra at fixed temperature ($T$\/=100 K, Fig. \ref{Fig4}) 
shows clearly the structural changes upon variation of the
oxygen concentration are most pronounced at $R\geq 3.5$ $\rm\AA$, 
a regime strongly dominated by MS. Thus extraction of 
meaningful structural information from our data requires MS to
be included in the data analysis.

\subsubsection{\label{SubsubSS}Single Scattering}

The interference function $\chi (k)$ for a single back-scattering
(SS) configuration ($\Theta=180^{\circ}$) of $l=1$ photoelectrons
($K$\/-absorption) can be expressed as

\begin{equation}\label{interference}
k\chi(k)=S_0^2F(k)\int dr \frac{\rho(R)}{R^2}\exp(-2R/\lambda(k))
\times\sin\left[(2kR+\varphi(k)+2\delta(k)\right]
\end{equation}

\noindent
Here $k$ is the momentum of the photoelectron, $S_0^2$ is an
amplitude reduction factor to account for intrinsic losses
at the absorbing atom,
$F(k)$ the scattering amplitude, $\rho(R)$ is a radial pair
distribution function (RDF), $\varphi(k)$ the scattering phase
shift, $\delta(k)$ the final-state phase shift at the absorbing
(central) atom. $\lambda (k)$ refers to the decay of the
photoelectron wave-function amplitude due to various losses.
These losses are taken into account by the imaginary parts of
the photoelectron self-energy and the core-hole width. 

$\rho(R)dr$ is proportional to the probability of finding an
atom of a given shell within $R+dR$ around the absorbing atom,
and $\int\rho(R)dr=N$ defines the number of atoms, $N$, 
in this shell. From Eqn. \ref{interference} $\chi k$ 
is the Fourier transform of an effective RDF, which includes the
structural information and the physics of the electron
scattering. The $true$ RDF can be expressed in moments of the
true RDF writing Eqn. \ref{interference} as

\begin{equation}\label{kchi}
k\chi (k)=A(k)\sin\Phi(k)
\end{equation}

\noindent
where

\begin{equation}\label{amplitude}
A(k)=\frac{S_0^2F(k)}{R^2}N\exp\left[-2R/\lambda(k)\right]
\exp(-2\sigma^2k^2)
\end{equation}

\noindent
and

\begin{equation}\label{phase}
\Phi(k)=2k
\left[R-\sigma^2\left(\frac{2}{R}+
\frac{2}{\lambda(k)}\right)\right]
-\frac{4}{3}c_3k^3+\phi(k).
\end{equation}

\noindent
$\sigma^2=\langle (R-R_0)^2\rangle$ 
denotes the mean-squared relative displacements from small
harmonic disorder of thermal and/or static origin. $c_3$ denotes
the mean-cubic relative displacement describing deviations from
the Gaussian RDF. Extraction of $c_3$ from the sine argument in
Eqn. \ref{kchi} is particularly useful in measuring the strength
of anharmonicities and other non-Gaussian disorder as a function
of temperature. The bond lengths and $\sigma_0^2$ listed in Tabs.
\ref{YO23}, \ref{YCu2} were determined by least-squares fitting
of the experimental $k$\/-space and/or $R$\/-space data to Eqn. 
\ref{interference}. $\Delta\sigma^2(T,x)$ was determined from
polynominal fits to the $\ln$\/-ratios of the envelopes

\begin{equation}\label{logratio}
\ln\left[\frac{A}{A_0}\right]=-2\Delta\sigma^2 k^2.
\end{equation}

\noindent
Assuming harmonic motion for a diatomic system the vibrational
contribution to the disorder is given by

\begin{equation}\label{Einstein}
\sigma_{vib}^2=\frac{h}{8\pi^2\mu\nu}\coth\frac{h\nu}{2k_BT}
\end{equation}

\noindent
where $\mu$ is the reduced mass, and $\nu$ the vibration
frequency.

$\Delta c_3(T,x)$ of the anharmonic Y--O2,3 RDF were derived
from polynominal fits to differences of Eqn. \ref{phase}. 

\begin{equation}\label{phasendiff}
\frac{1}{k}\left(\Phi-\Phi_0\right) = 
2\left[(R-R_0)-\Delta\sigma^2\left[\frac{2}{R}
+\frac{2}{\lambda(k)}\right]\right]-\frac{4}{3}\Delta c_3 k^2
\end{equation}

\noindent
To extract the $true$ RDF from the experimental
absorption-fine-structure (carried out for the Y--Cu2 pair at
$T$\/=20 K, $cf.$ sec. \ref{subsubYCu2}), $\sigma^2$ and $c_3$
were determined from least-squares fits and interpolated between
$k$\/=3.5 and 0 $\rm\AA^{-1}$. $P(R)$ was then obtained by taking
the inverse sine transform from $k_{min}$\/=0 to $k_{max}$\/=12
$\rm\AA^{-1}$. For details of the ``splicing'' see $e.g.$  Ref.
\cite{SteBou}.

\subsubsection{\label{MS}Multiple Scattering}  

The MS contributions to $\chi k$ were
calculated using the high-order MS approach of the FEFF 6 code.
The theoretical ingredients of this code can be found in Ref.
\cite{ZabEll}. The multiple scattering paths are expressed in a
form analogous to single back-scattering given in Eqn.
\ref{interference}, but which includes all multiple scattering
and curved wave effects. For MS configurations the effective
path length is given by $R=r_{tot}/2$, where $r_{tot}$ is the
total length of the MS path. The geometrical data of the
relevant MS configurations are listed in Tabs. 
\ref{YO23}, \ref{YCu2}, \ref{Y62}, \ref{Y5}, \ref{Y35}. Calculated
Fourier transform spectra are displayed in Fig. \ref{Fig17}.
From a total of 598 relevant scattering configurations,  found
in a $R$\/=8 $\rm\AA$ sized cluster (choosing the default curved
wave and plane wave filters of the code), only 33 (1-4 legs)
turned out to reproduce qualitatively the experimentally
observed spectra. 

Application of more elaborate procedures yielding quantitative
results from the high\/-$R$ scattering configurations first of all
needs a realistic model, which contains 
the ''allowed`` distortions of the important MS configurations.
 In the following
sections we shall develop step by step such a model, which
in particular describes the static atomic displacements 
that are at the origin of the alterations of the strongest MS
observed on doping and variation of temperature ($T<100$ K). 
This model might serve as a
starting point for a future quantitative analysis of the
complete data sets.

The variations in the peak heights of peaks ''6.2`` and ''3.5``
are found to depend sensitively on the position of the Cu2 atom
in a few MS paths. We have determined quantitatively the 
correlations between peak heights and Cu2 displacements 
(listed in Tab. \ref{MSCu}) from re-iterated full calculations
of the MS within the whole cluster, and by displacing the Cu2 atoms 
from their nominal position along the $c$\/-axis. The mean-squared
variations in $R$, the so-called path Debye-Waller factors,
$\sigma_{\Gamma}^{2}$, have been chosen to be zero for all paths.
The Gaussian broadening in the calculated Fourier transform
(Fig. \ref{Fig17}) spectra therefore arises solely from the 
finite transform range 
(2-16 $\rm\AA^{-1}$) and the Gaussian window function (0.2
$\rm\AA^{-1}$).

\subsubsection{\label{subsubCentral}The Atomic Cluster Centered
at the yttrium Site}

\sfig{Fig2} 
{Distribution of the relative amplitudes of the Y $K$\/-EXAFS as
calculated from a cluster of 149 atoms ($R$\/=8 $\rm\AA$) in
$\rm YBa_2Cu_3O_{6.9}$ using the FEFF6 code and
cry\-stal\-logra\-phic data. The $k$\/-range is
0-20 $\rm\AA^{-1}$. Amplitudes are given relative to the
''strongest`` amplitude ($\equiv$ O2). Plotted are 598 bars
corresponding to the filtered scattering configurations out of a
total of $\simeq 8500$ with up to 8 legs.}  
{12.5}     
{7.88}   
{7.726}   
{4.}      
{t}        

The photoexcited yttrium atom in $\rm YBa_2Cu_3O_x$ is an
ideal observer of the atomic structure of the $\rm
CuO_2$\/-planes. Yttrium is located in the electronically inert
''separating`` layer in between the $\rm CuO_2$ double layers,
see $e.g$ Ref. \cite{ShaJor}. Its position and the vibrational
dynamics are expected to be only weakly affected by the oxygen
vacancies at the chain sites and the related order--disorder
phenomena. Nearest neighbours of Y 
are the planar oxygen atoms, O2 and O3, next nearest neighbours the
planar copper atoms, Cu2. Their RDF's are expected to show up in
the Fourier transform spectra well isolated from each other.
Thus Y-EXAFS yields directly information on the dimples in the
$\rm CuO_2$ plane. From the crystalline symmetry we expect the
Y--Cu2 pair to be unaffected by the orthorhombic distortion of
the unit cell.

\nfig{Fig3}                                        
{Yttrium $K$\/ EXAFS in $\rm YBa_2Cu_3O_{x}$ as a function of
oxygen concentration at $T$\/=100 K: co-plotted photoelectron
interference patterns, $\chi k^2$, from the four oxygen
concentrations under investigation.}    
{11.424} 
{7.8625}  
{12.5}   
{t}      

The selection of an ``inert'' central site is crucial for
EXAFS measurements of subtle local distortions. For instance
Cu-EXAFS in $\rm YBa_2Cu_3O_x$ suffers from the two
nonequivalent sites of copper. Ba might be considered as an
equally well suited candidate for the central site 
as Y. However, the position of Ba
has been shown to be displaced by oxygen vacancies in the chains.
Diffraction studies of the ortho-II superstructure in underdoped
$\rm YBa_2Cu_3O_x$ ($x$\/=6.40-6.7) report shifts by +0.04
$\rm\AA$ along the $a$\/-axis \cite{ZeiHoh}. Thus the
$\{100\}$-mirror plane through the average barium sites is
partially removed, at least in underdoped $\rm
YBa_2Cu_3O_{7-\delta}$. Much weaker antiferro-type shifts of
$\simeq 0.011$ $\rm\AA$ have been reported for the Y atoms.

Fig. \ref{Fig1} displays a view of a Y--O2,3--Cu--Ba cluster
with some of the important scattering paths contributing to the
Y EXAFS (arrowed):

\begin{itemize}
\item Y--O2, Y--O3 (SS), see Tab. \ref{YO23}
\item Y--Cu2 (SS), see Tab. \ref{YCu2}
\item Y--Cu2--Ba (SS, MS), see Tab. \ref{Y62}
\item Y--O2--Ba, Y--O3--Ba  (SS, MS), see Tab. \ref{Y5}
\item Y--O2--Cu2, Y--O3--Cu2 (MS), see Tab. \ref{Y35}
\item Y--O2--O3 (MS), see Tab. \ref{Y35}
\end{itemize}

\subsection{\label{subsample}SAMPLE PREPARATION}
\nfig{Fig4}                              
{Yttrium $K$\/ EXAFS in $\rm YBa_2Cu_3O_x$ as a function of
oxygen concentration at $T$\/=100 K: co-plotted moduli of the 
Fourier transforms, $\mid {\rm FT}(\chi k^2)\mid$, as obtained 
from the interference patterns in Fig. \ref{Fig3} using 
$k$\/=2--16 $\rm\AA^{-1}$. Note the reduced height of peak
''6.2`` for $x$\/=6.801.}   
{11.844}  
{7.9645} 
{12.5}  
{b}     

\noindent
The samples under investigation were polycrystalline powders
with oxygen contents: $x=6.806$
(``underdoped''), $x=6.947$ (``optimum doped''), $x=6.968$
(``overdoped''), $x=6.984$ (``heavily overdoped''). The
samples were from the same batches earlier used for the 
measurements of the 
crystallographic, electrical and magnetic properties 
in the underdoped--overdoped transition \cite{ConJil}. 
We used preparations which were obtained by direct
reaction of $\rm BaCO_3$ with the metal oxides. The reduction
was performed in equilibrium with $\rm YBa_2Cu_3O_{6.05}$ in
sealed ampoules (CAR method). For further details of the preparation
route and the precision measurement of the oxygen contents see
Ref. \cite{ConJil} and references therein. 
\nfig{Fig5}                                        
{Yttrium $K$\/ EXAFS in $\rm YBa_2Cu_3O_{6.968}$ as a function of
temperature. 18 photoelectron interference patterns, 
$\chi k^2$, at $T$=86--250 K are co-plotted.}    
{11.696} 
{7.8625}  
{12.5}   
{t}      
The x-ray absorbers were prepared from 
about 60 mg of finely ground powders ($\leq 5\mu$m) 
spread onto a metallized Kapton tape and sealed by Kapton foil.

\subsection{\label{subAbs}ABSORPTION SPECTROSCOPY}

A stack of 8 such foils was found to yield the optimum
absorption contrast, $\Delta\mu d\simeq 0.69\pm 0.03$. The x-ray
absorption measurements were carried out in transmission mode
using the double crystal spectrometer at the bending magnet
BM29(BL18) at the European Synchrotron Radiation Facility
(ESRF). The harmonics were suppressed to $\leq 10^{-3}$ by
detuning the flat Si 111 crystals to 50\% of the maximum of the
rocking curve. Krypton filled ionization chambers served as
detectors. The spectrometer broadening was set to match the Y
$K$\/-core-level width. The scans were extended up to 1500 eV
above the Y $K$\/-edge at 17033 eV, yielding 
$\simeq 20$ $\rm\AA^{-1}$. $E_0$ was uniquely defined by the 
inflection point at the onset
of the white line. The drifts in energy caused by the thermal
load on the first uncooled monochromator crystal 
were typically $\pm 1.5$ eV.

The low temperature experiments were carried out using 
a closed cycle He-cryostat. 
The absorption foils were coupled to the cold
head by He gas. The temperature drifts during high temperature
scans were $<\pm 0.1$ K, during measurements across the
superconducting phase transition $<0.05$ K. We took care of
possible hysteretic effects during the thermal cycle. Each
sample was cooled down within about 2 h from room temperature
to about 18 K, subsequently heated up to room temperature, and again
cooled down to 18 K. Possible hysteretic effects turned out to
be within the error margins of the data reduction procedure. All
data presented in the following sections were recorded at
increasing temperature. The time intervals between successive
temperature scans were in average about 1 h.

\nfig{Fig6}                              
{Yttrium $K$\/ EXAFS in $\rm YBa_2Cu_3O_{6.968}$ as a function of
temperature, T=86--250 K: co-plotted moduli of the 
Fouriertransforms, $\mid {\rm FT}(\chi k^2)\mid$, from the
interference patterns in Fig. \ref{Fig5}. Transform range
$k$\/=2--16 $\rm\AA^{-1}$.}   
{11.424}  
{7.888} 
{12.5}  
{b}     

\section{\label{secExpRes}Experimental Results}
Figs. \ref{Fig3}, \ref{Fig5} display typical photoelectron
interference patterns, $\chi k^2$, as obtained by data reduction
of the raw absorption spectra, $\mu d(E)$, where $\mu $ is the
linear absorption coefficient, $d$ the thickness of the absorber
and $E$ the photon energy. After subtraction of a linear
pre-edge background the spectra were normalized and the atomic
absorption,\/ $\mu_0$, was determined from a sequence of four cubic
splines. The energy dependence of $\mu_0$ above threshold was taken
into account using tabulated numbers. No further corrections to the
data were applied, neither to eliminate possible atomic multielectron
excitations, nor to smooth the monochromator glitch occurring
systematically around $k$=8 $\rm\AA^{-1}$. The accuracy and
reproducibility of the data reduction procedure can be appreciated
from $\chi(k) k^2$ (Figs. \ref{Fig3},\ref{Fig5}), from the
corresponding Fourier transform spectra (Figs. \ref{Fig4},\ref{Fig6}),
and in particular from the behaviour of $\mid {\rm FT}(\chi k^2)\mid$
for $R_{eff}\rightarrow 0$. The tiny bump around 1 $\rm\AA$ is an
artefact due to deficiencies in the above background definition.

All spectra were recorded up to $k$\/=20 $\rm\AA^{-1}$, but only the
window $k$\/=2--16 $\rm\AA^{-1}$ (Gaussian broadened by 0.2
$\rm\AA^{-1}$) was used for the structural analysis. Some of the
spectra exhibit spikes at $k\ge 16$ $\rm\AA^{-1}$ (Fig. \ref{Fig3})
degrading the important nearest neighbour Y--O2,3 signal. Scans with
an excellent $S/N$ ratio were obtained from the ``overdoped''
(x=6.968) sample, in particular during one synchrotron shift at
$86>T>105$ K. We have used this set of data to study in greater detail
the subtle structural anomalies showing up in the direct vicinity of
$T_c$ (Fig. \ref{YCuSig}).

Within $R_{eff}$=0-8 $\rm\AA$ the Fourier transform spectra, $\mid
{\rm FT}(\chi k^2)\mid$, exhibit 6 prominent peaks. The calculated
distribution of scattering amplitudes displayed in Fig. \ref{Fig2}
exhibits a similar overall pattern. Single scattering at the oxygen
and copper atoms in the $\rm CuO_2$ plane gives rise to the peaks
``O2,3'' and ``Cu2''; at larger $R_{eff}$ mixtures of single and
multiple scattering configurations cause the peaks labelled ``3.5'',
``4'', ``5'', ``6.2''. Tabs. \ref{YO23}, \ref{YCu2}, \ref{Y35},
\ref{Y5}, \ref{Y62} list the strongest scattering configurations,
their geometries and degeneracies, the atomic coordinates, and the
relative amplitudes as calculated from an average of the unit cell
parameters \cite{FraFis,SchErb,CapTou}. The overall patterns of the
Fourier transform spectra are nearly identical for all concentrations
at 100 K (Fig. \ref{Fig4}). Differences occur in the peak heights, but
not in the peak positions. It is important to note that peak ``6.2''
in the ``underdoped'' sample ($x$=6.801, short dashed line) is
strongly damped. A weaker but clear reduction is observable in peak
``3.5''. Seemingly peak ``5'' is not affected by the oxygen
concentration. As a function of temperature, peak ``6.2'' exhibits the
strongest relative variation (Fig. \ref{Fig6}) although the
temperature also affects the height of peak ``5''.

We first analyze the single scattering (SS) paths from the
nearest and next nearest neighbours Y--O2,3 and Y--Cu2 pairs in
terms of bond lengths, $R$, mean-squared radial deviations,
$\sigma ^2$, and characteristic vibrational frequencies,
${\nu}$. Then we study the SS and MS paths contributing
to the three high-$R$ peaks ``3.5'', ``5'', ``6.2'', and 
show their dependence on $x$ and $T$. Finally we discuss
the evolution of the static disorder in the
Y--Cu2\---O2\---O3\---Ba cluster with $x$ and $T$. In particular
we focus on the three-body correlations Y--Cu2\---Ba (``6.2''),
Y--O2,3\---Ba (``5''), Y--Cu2\---O2,3, and Y--O2,3\---O2,3
(``3.5''), which exhibit significant variations at $T<T_c$. The
relatively weak peak ``4'' here will not be addressed further. 
\begin{table}[b]
\begin{center}
\caption{\label{YO23}Y--O2,O3 single scattering configurations
as expected from the crystallographic unit cell. Atomic
coordinates are in $\rm\AA$. $R_{avg}$ is half of the total
scattering length. Scattering angles, $\Theta$, are 
180$\rm^{\circ}$ for back-scattering, 0$\rm^{\circ}$ for
forward-scattering, {\it g} the degeneracy of the scattering
configurations. Rel. Amp. denotes the relative EXAFS amplitude
(see text). \# is a counter.}
\begin{tabular}{cccccrccc} 
\hline
$R_{avg}$[\AA]&scatterer&\it x&\it y&\it z&$ \Theta$ [$^{\circ}$]&
\it g&Rel. Amp. [\%]&\#\\ 
\hline
2.377&O2&-1.911&0&1.413&180.00&4&100.0&1\\
 &Y&0&0&0&180.00& & &\\
\\
2.416&O3&0&-1.943&1.436&180.00&4&95.6&2\\
 &Y&0&0&0&180.00& & &\\ 
\hline
\end{tabular}
\end{center}
\end{table}

\subsection{\label{subNext}NEXT AND NEXT NEAREST NEIGHBOURS}

The $\rm CuO_2$\/-planes in $\rm YBa_2Cu_3O_{7-\delta}$ are not
flat but dimpled. In contrast to the buckled $\rm CuO_2$ planes
in the 214 compounds the planar oxygen atoms in $\rm
YBa_2Cu_3O_x$ are all displaced from the Cu2 planes in the same
direction. Therefore the geometry of the $\rm CuO_2$ planes in $\rm
YBa_2Cu_3O_x$ may be seen as stacked O2,3 and
Cu2 layers separated by about 0.27 $\rm\AA$ in $c$\/ direction.
The O2,3 layers are located closest to the Y layer, and the Cu2
layers closest to the Ba layer.

\nfig{Fig7}                                        
{Y--O2,3 mean-squared displacements, $\sigma^2$, as a function
of temperature. $Left:$\/''Underdoped`` and ''heavily
 overdoped`` samples, $Right:$\/''optimum doped`` (closed
circles) and ''overdoped samples`` (open circles) of $\rm
YBa_2Cu_3O_x$. The solid lines are fits from the harmonic model
of lattice vibrations.}    
{12.09} 
{6.264}  
{12.5}   
{t}      

\nfig{Fig8}                                        
{Y--O2,3 mean-cubic relative displacements, $\Delta c_3$, 
as a function of temperature for $x$\/=6.968, relative to
$T$\/=100 K. Fits were from 7-10 $\rm\AA^{-1}$. 
Lines connecting the data points are guides to the
eyes. $Right:$\/ Zoom of the region around $T_c$.}    
{12.28} 
{6.03}  
{12.5}  
{b}     

The spacing between the O2,3 and Cu2 layers gives rise to a
difference of about 0.8 $\rm\AA$ between the average Y-O2,3 and
Y-Cu2 bond lengths. The Fourier transform spectra of high resolution
Y EXAFS ought to exhibit well isolated O2,3 nearest neighbours 
from the Cu2 next nearest neighbour shell. Since the
nearest high order shells are about +0.6 $\rm\AA$ further out 
than the Cu2 shell ($cf.$ Tab. \ref{Y35}), both, the Y--O2,3 and
the Y--Cu2 single scattering signals are well isolated from
the other scattering contributions, see Figs. \ref{Fig4},\ref{Fig6}.
Standard single scattering analysis is therefore expected to
yield straightforward structural information on the O2,3 and Cu2
layers and their interlayer spacing.

\subsubsection{\label{subsubYO23}The Y--O2,3 Pair}

The orthorhombic nature of superconducting $\rm YBa_2Cu_3O_x$ is
manifest in the oxygen layer of the $\rm CuO_2$ plane by the two
symmetry equivalent sites O2 and O3, the differences in their
thermal vibrations and the differences in their displacements
from the Cu2 layer, see $e.g.$ Ref. \cite{FraFis}. 
The two displacements have been reported to exhibit 
magnitudes differing by $<0.01$ $\rm\AA$, but other 
diffraction studies have reported flat O2,3 layers.  

The average Y--O2,3 bond lengths are 2.40 $\rm\AA$, 
and depend weakly on temperature (4--320 K), see $e.g$ Ref.
\cite{FraFis,SchErb,CapTou,JorKwo}, 
but strongly on oxygen concentration.
Variation of $x$ from 6.6 to 6.97 decreases 
the average Y--O2,3 bond length by 0.04 $\rm\AA$ (at 5
K) \cite{ConJil}. Y--O2 differs from Y--O3 by about 0.02 $\rm\AA$.
\begin{table}[htb]
\begin{center}
\caption{\label{O23Fit}Results obtaind from a harmonic single
shell fit to the average of the Y--O2,3 pairs.}
\begin{tabular}{ccccc} 
\hline
\it x&{$R$ [$\rm\AA$]}&{$\sigma_0^2$ (40 K) [$\rm\AA^2$]}&
$\bar\nu$ [$\rm cm^{-1}$]&$E_0$ [eV]\\
\hline
6.984&3.39(4)&0.0035(2)&253(25)&3(3)\\
6.968&3.39(4)&0.0051(2)&313(25)&3(3)\\
6.947&3.39(4)&0.0048(1)&291(25)&3(3)\\
6.806&3.39(4)&0.0048(2)&270(25)&3(3)\\
\hline
\end{tabular}
\end{center}
\end{table}
Fits of a single shell absorption fine structure to the filtered
($R_{eff}$\/=1.1--2.2 $\rm\AA$) Y--O2,3 shell at $T < T_c$ (40 K)
and at $T > T_c$ (100 K) yielded only coarse agreement with the
experimental data, even when narrowing the $k$\/-range to 4--9
$\rm\AA^{-1}$. Inclusion of the orthorhombic splitting (0.02
$\rm\AA$) did not improve these fits, 
even when the splitting was increased up
to 0.2 $\rm\AA$. Obviously the Y--O2,3 pairs are strongly
disordered and/or vibrate strongly anharmonically. 

The presence of disorder or strong anharmonicities in the
Y--O2,3 pairs is directly visible in the imaginary part of the
Fourier transform spectra (not shown here). 
Using the nominal energy zero, 
$E_0$, the maximum of Im$(\chi k^2)$ is found markedly shifted 
apart the maximum of the modulus, and even the best value of 
$E_0=+3\pm 0.3$ eV (resulting from the harmonic 
$k$\/-space fits) does not bring them into
coincidence. Surprisingly, the average Y--O2,3 bond lengths 
obtained from these fits (see Tab. \ref{O23Fit}) 
are found close to those reported from the diffraction studies. 
Clearly, applying the harmonic approximation to an anharmonic
problem we obtain only 
approximate parameters, since we put a sufficiently broadened 
Gaussian RDF on an anharmonic RDF, at best at its centre of gravity. 

We have quantified the anharmonicity of the Y--O2,3 RDF's by the
relative mean-cubic relative deviation, $\Delta c_3$, 
extracted from the $T$\/-dependent differences 
of the sine arguments, $\Delta\Phi$. In our best set of data 
($x$\/=6.968) we find a marked singularity of
$\Delta c_3(T)$ exactly at $T_c$ (Fig. \ref{Fig8}). 
\nfig{Fig9} 
{Y--Cu2 mean-squared displacements, $\sigma^2$, as a function of
temperature. Determined from the $\ln$\/-ratios. 
$Left:$\/ The solid line is a fit from the harmonic model of 
lattice vibrations yielding $\bar\nu=192$
 cm$^{-1}$. $Right:$\/ Same data but zoomed in the superconducting
phase. Line connecting the data points is a 
guide to the eyes.}       
{12.48}    
{5.86}     
{12.5}     
{t}        
The relative dependence of disorder on the oxygen 
concentration at 40 K, expressed as 
$\sigma_0^2(x)$\/-$\sigma_0^2(x=6.947)$, was determined from 
the $\ln$\/-ratios (Eqn. \ref{logratio}) at $k^2$\/=16-81 
$\rm\AA^{-2}$. $\sigma_0^2$ is independent of the oxygen 
concentration for $x$\/=6.801-6.947, but decreases by about 20\% 
in the ''heavily overdoped`` sample ($x$\/=6.984). 
This finding corroborates the relative heights of peak ''5``, 
$cf.$ Fig. \ref{Fig14}. 

The temperature dependencies of the mean-squared-deviations,
$\sigma^2(T)$, were determined based on the assumption that
$\sigma^2$ obeys a harmonic model of vibrations (Eqn. \ref{Einstein}). 
The results are displayed in Fig. \ref{Fig7} (drawn out lines) 
and listed in Tab. \ref{O23Fit}. 
Although the cha\-rac\-teristic frequencies, $\nu$, 
are only coarse approximations, they indicate at least tendencies in 
the vibrational dynamics on doping. We 
note that the Y--O2,3 vibrations harden as the
oxygen concentration increases up to $x$\/=6.968, but soften for
$x$\/=6.984. As expected $\bar\nu=\nu/c$  
is different, but not far from the wavenumber of the
infrared active mode reported to be at 312 $\rm cm^{-1}$ 
for the ''oxygen O2,3 mode`` \cite{HomHar}.

\subsubsection{\label{subsubYCu2}The Y--Cu2 Pair}

The Y--Cu2 peak appears in the Fourier transform spectra well
isolated from its neigbors. From the crystallographic structure
the Y--Cu2 bond lengths are $\simeq 2.2$ $\rm\AA$ and are not
orthorhombically split. Harmonic single shell fits were
performed in $k$\/-space (5-12 $\rm\AA^{-1}$) and the results 
for 40 K and 100 K are listed in Tab. \ref{YCu2Fit}. 
\begin{table}[t]
\begin{center}
\caption{\label{YCu2Fit}Results from harmonic 
single shell fits to the Y--Cu2 peak.}
\begin{tabular}{ccccccc} 
\hline
\it x&\multicolumn{2}{c}{$R$ [$\rm\AA$]}&\multicolumn{2}{c}
{$\sigma^2$ [$\rm\AA^2$]}&
$\bar\nu$ [$\rm cm^{-1}$]&$E_0$ [eV]\\
&40 K&100 K&40 K&100K& & \\
\hline
6.984&3.203(5)&3.204(5)&0.0023(1)&0.0031(1)&192(25)&5.93(19)\\
6.968&3.204(5)&3.203(5)&0.0021(1)&0.0033(1)&192(25)&5.56(21)\\
6.947&3.200(5)&3.200(5)&0.0021(1)&0.0032(1)&192(25)&5.92(22)\\
6.806&3.200(5)&3.200(5)&0.0021(1)&0.0031(1)&192(25)&5.88(22)\\
\hline
\end{tabular}
\end{center}
\end{table}

For all concentrations we find within the error bars 
the same energy zero, $E_0$,\/ pointing to the same degree of
\begin{table}[b]
\begin{center}
\caption{\label{YCu2}Geometrical data of the Y--Cu2 single
scattering configuration giving rise to the isolated 'Cu2 peak'.
Calculated from crystallographic data. $R$ is half of the total
scattering length. $\Theta$ denotes the scattering angles
(180$\rm^{\circ}$ $\equiv$ back-scattering, 0$\rm^{\circ}$
$\equiv$ forward-scattering), {\it g} the degeneracy of the
scattering configurations. Rel. Amp. is the amplitude of the
scattering configuration (see text). \# is a counter.}
\begin{tabular}{cccccrccc} 
\hline
$R$[\AA]&scatterer&\it x&\it y&\it z&$ \Theta$ [$^{\circ}$]&
\it g&Rel. Amp. [\%]&\#\\ 
\hline
3.209&Cu2&1.911&1.943&1.682&180.00&8&96.2&1\\
 &Y&0&0&0&180.00& & &\\
\hline
\end{tabular}
\end{center}
\end{table}
disorder in the Y-Cu2 pairs, if one exists. 
Also the bond lengths $R_{Y-Cu2}$ are found to be independent of 
oxygen concentration and of temperature between 20 and 100 K. 
However, the temperature dependencies of the mean-squared
deviations, $\sigma ^2(T)$, exhibit a step-like behaviour around
80 K deviating from the harmonic contribution to the
Debye-Waller factor (solid line in Fig. \ref{Fig9}). We
find $\Delta\sigma^2_{100-40 K}\simeq 0.001$ $\rm\AA^{2}$ from
the $k$\/-space fits and $\simeq 0.0006$ $\rm\AA^{2}$ from the
$\ln$\/-ratio. 

Since the Y--Cu2 pair is well isolated from its neighboured
pairs, and nominally does not exhibit orthorhombic splitting, we
used the so-called splice-method to extract the $true$ pair 
distribution function, $P(R)$, from the
experimental data. Using theoretical phase shifts and envelope
functions from the FEFF code, we spliced the phase differences
and $\ln$\/-ratios of the envelope functions from 3.5 to 0
$\rm\AA^{-1}$. Fig. \ref{YSplice} exhibits for $x$\/=6.801 and 20 K 
the resulting $P(R)$ (thick solid line). We note a
weak asymmetry of $P(R)$ towards shorter bond lengths, $i.e.$ a
rather weak leakage into the O2,3 layer. For comparison we show an ideal
Gaussian (thin dotted line) and a spliced Gaussian (dashed
line). The former indicates the broadening introduced by the
finite $k$\/-range and the Gaussian window function. The latter
has been calculated from a theoretical and exponentially dampened
spectrum using the same splice as for the experimental data.
Thus we may estimate the error introduced by the interpolation
of the experimental data. We find the $true$ $P(R)$ only
slightly outside this margin of error, confirming the absence of
strong disorder and/or strong anharmonicities. 

We note that the absence of anomalous copper vibrations in $\rm
YBa_2Cu_3O_x$ ($x$=6.2, $x$=7) between 10 and 300 K has been
also concluded from neutron resonance absorption spectroscopy
\cite{HecGol}.

\sfig{Fig10}                                        
{Pair dis\-tri\-bu\-tion func\-tion, $P(R)$, of the Y--\-Cu2 pair in
the ''under\-doped`` sample ($x$\/=6.801) at 20 K (thick).
De\-ter\-mined from the ex\-pe\-ri\-men\-tal data using the
splice me\-thod. For com\-parison: an ideal Gaussian (thin dotted)
and a ''spliced`` Gaussian (dashed), see text.}    
{12.5}   
{8.704} 
{6.392}  
{3.5}     
{t}      

\subsection{\label{subPeaks}
PEAKS FROM MIXED SCATTERING CONFIGURATIONS}

The peaks at $R\geq 3.5$ $\rm\AA$ in the Fourier transform spectra
arise from mixed scattering configurations, $i.e$ 2-leg single
scattering paths, which are nearly degenerate with 3- and 4-leg
multiple scattering paths, and/or nearly degenerate single and
multiple scattering paths from different atomic configurations. Due to
the distinguished position of the central yttrium atom at the
inversion centre of the unit cell, and due to the particular
interlayer spacings in $\rm YBa_2Cu_3O_x$, the yttrium EXAFS yield
many nearly collinear scattering
\begin{table}[t]
\begin{center}
\caption{\label{Y62}
Geometrical data of the strongest scattering configurations
contributing to peak ``6.2''. Calculated from crystallographic
data. $R_{avg}$ is half of the total scattering
length. $\Theta$ denotes the scattering angles (180$\rm^{\circ}$
$\equiv$ back-scattering, 0$\rm^{\circ}$ $\equiv$
forward-scattering), {\it g} the degeneracy of the scattering
configurations. Rel. Amp. is the amplitude of the scattering
configuration (see text). \# is a counter}
\begin{tabular}{cccccrccc} 
\hline
$R_{avg}$[\AA]&scatterer&\it x&\it y&\it z&$ \Theta$ [$^{\circ}$]&
\it g&Rel. Amp. [\%]&\#\\ 
\hline
6.583&Ba&3.823&3.886&-3.691&180.00&8&17.5&1\\
     &Y&0&0&0&180.00& & &\\
\\
6.586&Ba&3.823&3.886&-3.691&177.70&16&34.4&2\\
     &Cu2&1.911&1.943&-1.682&4.71& & & \\
     &Y&0&0&0&177.57& & &\\ 
\\
6.589&Cu2&1.911&1.943&1.682&4.71&8&19.9&3\\
     &Ba&3.823&3.886&3.691&180.00& & & \\
     &Cu2&1.911&1.943&1.682&4.71& & & \\
     &Y&0&0&0&180.00& & &\\ 
\hline
\end{tabular}
\end{center}
\end{table}
configurations. We take advantage of some of them, in
particular of those exhibiting strong or very strong scattering
amplitudes. Yttrium in its photoexcited state interferes with
the copper and oxygen atoms in the $\rm CuO_2$ plane and through
three-body configurations with the Ba atoms in the doping block.
The shortest Y--O2,3--Ba and Y--Cu2--Ba scattering
configurations deviate from collinear ($0^{\circ}$) scattering
geometries only by about $14^{\circ}$ and $5^{\circ}$ and hence
give rise to the prominent peaks at 5 $\rm\AA$ and 6.2 $\rm\AA$,
respectively. Small deviations from collinearity have very strong
effects on the three-body scattering amplitude. The results from
our MS calculations with adjustable Cu2 position are listed in
Tab. \ref{MSCu}.

\subsubsection{\label{subsub62}Peak ``6.2''}

From the calculated distribution of scattering
amplitudes displayed in Fig. \ref{Fig2} we expect the 
spike at $R\simeq 6.58$ $\rm\AA$ to give rise to the peak ``6.2''. 
The weak side band at the low $R$ side has relative amplitudes
up to 13\%, and obviously causes the bump between the peaks
``5'' and ``6.2''. It will not be considered further. Table \ref{Y62} 
lists the scattering configurations contributing to
the ``6.2'' peak. We find three Y--Ba scattering configurations
with nearly degenerate scattering lengths centred at
$R_{avg}=6.586$\/: single back-scattering Y--Ba (\#1), and two
nearly collinear Y--Cu2--Ba MS configurations (\#2,\#3), probing
a cluster of $3\times 3$ unit cells. Like
the Y--Cu2 SS configuration (``Cu2'') these three configurations
are not orthorhombically split. The 3-leg MS configuration (\#2)
contributes nominally by about 47\%, the 4-leg configuration by
about 27\% to the total scattering amplitude at ''6.2'``
(neglecting the $\simeq 8\%$ constant MS background in this
region). However it is important to note that the high sensitivity for
the detection of subtle Cu2 displacements from the collinear 
three-body scattering configuration arises from the phase 
contrast between the 3-leg and 4-leg MS configurations
\cite{RoeCru}.

Fig. \ref{YCuBa} displays the concentration and temperature
dependence of the maxima of $\mid {\rm FT}(\chi k^2)\mid$ at
$R_{eff}\simeq 6.2$ $\rm\AA$. Within the scatter of the data 
the peak heights from the ``optimum doped'' and both
overdoped samples (closed symbols) exhibit congruent temperature
dependences at $\simeq 70-300$ K, and saturation for
$T\rightarrow 20$ K. Seemingly there is a curvature at high
temperatures, which deviates from the inverted $\cosh (1/T)$ (see
Eqn. \ref{Einstein}) behaviour expected for 
harmonic Y--Ba vibrations. The saturation
behaviour of the peak heights in the superconducting phase is
different for the ``heavily overdoped'' and the two others. The
former starts to saturate at $T\leq 40$ K, whereas the latter two
saturate at $T\leq 60$ K and level out at values
$\simeq 10$\% lower, see Fig. \ref{YCuBasc}.

The peak heights of the ``underdoped'' sample (open rectangular
symbols) have a significantly weaker temperature dependence but
exhibit distinct ``staircases'' at $\simeq 160$ K 
and $\simeq 100$ K. At
$T>250$ K we find that the temperature dependencies for all
concentrations converge. In the superconducting phase the
peak height is relatively decreased by $\simeq 15\%$, and a
broad maximum seems to evolve around $\simeq 40$ K.

\sfig{Fig11} 
{Modulus of the Fourier transform 
at the ma\-ximum of the ``6.2'' peak as a function of 
temperature and concentration. For details see text. 
Lines connecting the data points are guides 
to the eyes.}    
{12.5}     
{8.120}    
{6.0502}   
{3.8}      
{t}        

\sfig{Fig12} 
{Modulus of the Fourier transform at the maximum of the ``6.2''
peak as a function of concentration at $T<120$ K. For details
see text. Lines connecting the data points are guides to the eyes. 
(Zoom of Fig. \ref{Fig11})}    
{12.5}     
{8.112}    
{6.4558}   
{3.8}      
{b}        

\subsubsection{\label{subsub5}Peak ``5''}

Peak ``5'' arises from similar Y-Ba scattering configurations as
peak ``6.2''. But the species of the intervening atoms
are orthorhombically split oxygens, O2 and O3, the 
average scattering length is shorter ($\simeq 5.354$ $\rm\AA$), 
and the bridging angle of $\simeq 15^\circ$ is larger compared to
$\simeq 5^\circ$. Fig. \ref{Fig2} shows two closely neighboured
spikes sitting in a narrow distribution of scattering
amplitudes. Tab. \ref{Y5} lists the geometrical data of the
\begin{table}[b]
\begin{center}
\caption{\label{Y5}Geometrical data of the
the strongest scattering configurations
contributing to peak ``5''. Calculated from 
crystallographic data. 
$R_{avg}$ is half of the total scattering length.
$\Theta$ denotes the scattering angles (180$\rm^{\circ}$ $\equiv$
back-scattering, 0$\rm^{\circ}$ $\equiv$ forward-scattering), {\it
g} the degeneracy of the scattering configurations. Rel. Amp. is
the amplitude of the scattering configuration (see text). \# is a
counter}

\begin{tabular}{cccccrccc} 
\hline
$R_{avg}$[\AA]&scatterer&\it x&\it y&\it z&$ \Theta$ [$^{\circ}$]&
\it g&Rel. Amp. [\%]&\#\\ 
\hline
5.314&Ba&-3.823&0&-3.691&180.00&4&17.7&1\\
     &Y&0&0&0&180.00& & &\\
\\
5.332&Ba&0&3.886&-3.691&173.99&8&31.7&2\\
     &O2&-1.911&0&-1.413&13.51& & & \\
     &Y&0&0&0&172.48& & &\\ 
\\
5.350&O2&1.911&1.682&1.413&13.51&4&14.5&3\\
     &Ba&-3.823&0&3.691&180.00& & & \\
     &O2&-1.911&0&1.413&13.51& & & \\
     &Y&0&0&0&180.00& & &\\ 
\\
5.359&Ba&0&3.886&-3.691&180.00&8&17.28&4\\
     &Y&0&0&0&180.00& & &\\
\\
5.376&Ba&0&3.886&-3.691&174.28&8&31.1&5\\
     &O3&0&1.943&-1.436&12.76& & & \\
     &Y&0&0&0&172.95& & &\\ 
\\
5.392&O3&0&-1.943&-1.436&12.76&4&14.3&6\\
     &Ba&0&-3.886&-3.691&180.00& & & \\
     &O3&0&-1.943&-1.436&12.76& & & \\
     &Y&0&0&0&180.00& & &\\
\\
5.451&Y&3.823&-3.886&0&180.00&4&12.9&7\\
     &Y&0&0&0&180.00& & &\\
\hline
\end{tabular}
\end{center}
\end{table}
strongest scattering configurations contributing to the peak
``5''. Due to the orthorhombic splitting we expect two triplets
(\#1-3, \#4-6), each with one Y--Ba SS configuration and two
Y--O--Ba MS configurations. \#7 labels a Y--Y single scattering
path. Since this particular Y--Y path is not orthorhombically
split and comparatively weak, we simply consider it as a 
constant contribution to the background.

\sfig{Fig13} 
{Modulus of the Fourier transform at the maximum of peak
``5'' as a function of oxygen concentration and temperature.
For details see text. Lines connecting the data points are
guides to the eyes.}    
{12.5}     
{8.112}    
{6.591}    
{3.8}      
{t}        

\sfig{Fig14} 
{Modulus of the Fourier transform at the maximum of peak
''5`` as a function of oxygen concentration at $T\leq 120$ K.
For details see text. Lines connecting the data points are
guides to the eyes. Zoomed from Fig. \ref{Fig13}}    
{12.5}     
{8.112}   
{6.4389}   
{3.8}      
{b}        

Due to the relatively large bridging angle of 
$\simeq 14^\circ$ we expect the 
sensitivity of the peak height to displacements of the
intervening oxygens to be relatively decreased. However, model 
calculations (not shown here) 
yielded clearly resolved variations of the peak 
height by $\geq 4$ \% for oxygen displacements of 
$\geq 0.01$ $\rm\AA$ in $c$ direction.        

For all oxygen concentrations under investigation the heights of
peak ``5'' are found to collapse into a single line (Fig.
\ref{Fig13}) for temperatures $80\leq T\leq 300$ K (within the
scatter of the data points). Seemingly in the normal phase 
 and close to $T_c$ the average position of the oxygen layer in the $\rm
CuO_2$ plane is not appreciably altered by variations of the
oxygen concentration from $x$\/=6.801-6.984. Furthermore
(within the limited accuracy of our analysis) the characteristic
vibrational frequencies of the Y--Ba pairs do not depend on
doping either.

\begin{table}[b]
\begin{center}
\caption{\label{Y35}Geometrical data of the
the scattering configurations
contributing to peak ``3.5''. Calculated 
from crystallographic data. 
$R_{avg}$ is half of the total scattering length.
$\Theta$ denotes the scattering angles (180$\rm^{\circ}$ $\equiv$
back-scattering, 0$\rm^{\circ}$ $\equiv$ forward-scattering), {\it
g} the degeneracy of the scattering configurations. Rel. Amp. is
the amplitude of the scattering configuration. \# is a
counter}
\begin{tabular}{cccccrccc} 
\hline
$R_{avg}$[\AA]&scatterer&\it x&\it y&\it z&$ \Theta$ [$^{\circ}$]&
\it g&Rel. Amp. [\%]&\#\\ 
\hline
3.691&Ba&0&0&-3.691&180.00&2&26.1&1\\
     &Y&0&0&0&180.00& & &\\
\\
3.759&O3&0&-1.943&1.436&125.32&16&17.2&2\\
     &O2&-1.911&0&1.413&123.97& & & \\
     &Y&0&0&0&110.70& & &\\ 
\\
3.770&Cu2&-1.911&-1.943&1.682&132.28&16&13.6&3\\
     &O2&-1.911&0&1.413&85.32& & & \\
     &Y&0&0&0&142.38& & &\\
\\
3.773&Cu2&1.911&-1.943&-1.682&131.20&16&13.7&4\\
     &O3&0&-1.943&-1.436&85.65& & & \\
     &Y&0&0&0&143.13& & &\\  
\\
3.823&Y&3.823&0&0&180.00&2&17.8&5\\
     &Y&0&0&0&180.00& & &\\
\\
3.886&Y&0&-3.886&0&180.00&2&17.0&6\\
     &Y&0&0&0&180.00& & &\\
\hline
\end{tabular}
\end{center}
\end{table}

Discrepancies up to $10\%$ in the heights of peak ``5'' are found
in the superconducting phases at $T<80$ K, between the ``heavily
overdoped'' sample on one hand, and the three others on
the other hand ($cf.$ Fig.
\ref{Fig14}).

\subsubsection{\label{subsub35}Peak ``3.5''}

A number of EXAFS works attribute peak ``3.5'' exclusively to Y--Ba
and Y--Y single scattering \cite{YanXu,BoyCla}. However, our multiple
scattering calculation finds three MS configurations to contribute
strongly to the total scattering amplitude at ``3.5'', nominally by up
to $\simeq 42$\% ! The geometrical data of the 6 scattering
configurations causing peak ``3.5'' are listed in Tab. \ref{Y35}. The
Y--Y single scattering is found to be strongest and orthorhombically
split (\#5, \#6), followed by the 
\sfig{Fig15} 
{Modulus of the Fourier transform at the maximum of peak
``3.5'' as a function of oxygen concentration and temperature.
For details see text. Lines connecting the data points are
guides to the eyes.  }    
{12.5}     
{8.112}    
{6.591}    
{3.8}      
{t}        
\sfig{Fig16} 
{Modulus of the Fourier transform at the maximum of peak
``3.5'' as a function of oxygen concentration at $T\leq 120$ K.
For details see text. Lines connecting the data points are
guides to the eyes. Zoomed from Fig. \ref{Fig15}.}    
{12.5}     
{8.6528}   
{6.2192}   
{3.8}      
{b}        
Y--Ba single scattering (\#1) perpendicular to the planes. 
The three triangular MS configurations:\/ Y--O2--O3 (\#2), Y--O2--Cu2
(\#3) and Y--O3--Cu2 (\#4) contribute significantly due to their high
degeneracies, $g=16$. It is important to note that these highly
degenerate MS configurations are nearly as sensitive to copper and
oxygen displacements in the $\rm CuO_2$ plane as the quasi collinear
MS configurations Y--O2,3--Ba (peak ``5''), and Y--Cu2--Ba (peak
``6.2''). It is clearly visible that the temperature and concentration
dependencies of the heights of peak ``3.5'' and peak ``6.2'' are
strongly correlated, in agreement with our model calculations
discussed further below.

Figs. \ref{Fig15}, \ref{Fig16} show the concentration and temperature
dependencies of peak ``3.5''. Both turn out to be very similar
to those of peak ``6.2'' ($cf.$ Figs. \ref{Fig11}, \ref{Fig12}), 
but not to those of peak ``5'' ($cf.$ Figs. \ref{Fig13}, \ref{Fig14}). 
The similarity includes the relatively weaker and staircase 
like temperature behaviour of the ``underdoped'' sample, and the 
congruent temperature behaviour of the op\-timum/\-over\-doped samples 
for $T>80$ K. It would thus appear that 
displacements of Cu2 are the common origin of the
similarities in Fig. \ref{Fig15} and Fig. \ref{Fig11}. 
However, there are also significant differences:  
$i.$ the heights of peak ``3.5''
exhibit a dip around 100 K, at least for the optimum/overdoped
samples, $ii.$ a plateau occurs between 100 K and 130 K, and $iii.$ 
at $T\leq 60$ K the order of peak heights changes. In Fig.
\ref{Fig12} x=6.968 has the weakest peak amplitude of the overdoped
samples, whereas in Fig. \ref{Fig14} it has the strongest.

In the superconducting phase, saturation behaviour evolves for the
optimum/overdoped samples (closed symbols); saturation behavior starts 
closest to $T_c$ for the ``optimum doped'' sample. The ``underdoped''
sample is found to exhibit still at 20 K a finite temperature
dependence of the peak height.

\section{\label{SecDisc}Discussion}
To summarize, we combine the structural results from the quantitative
analysis of the nearest and next nearest neighbours peaks, ``O2,3''
and ``Cu2'', with the qualitative information deduced from the
inspection of the mixed scattering configurations ``6.2'', ``5'', and
``3.5''. Here we restrict ourselves to the static changes of the
local geometry at low $T$\/, and to overall tendencies. 
A detailed analysis of the apparent and interesting anomalies 
occuring at $T_c$\/ and at other characteristic
temperatures has to include the vibrational dynamics of the many body
scattering configurations, and will be reported in forthcoming papers.

We start with the assumption that the position of the Cu2 atoms is the
key variable that controls the {\it relative} variations of the peak
heights at $R\geq 3.5$ $\rm\AA$ for the different oxygen
concentrations. To understand the effect of Cu2 displacements on the
Fourier transform spectra we carried out a series of model
calculations. We have been able to reproduce qualitatively the
experimental Fourier transform spectra and their variations with
doping at low temperatures. Although vibrational displacements were
completely neglected in these calculations ($\sigma^2$\/=0), we
believe that the results are indicative of the response of the Fourier
transform spectrum to such displacements. 

Some of the calculated Fourier transform spectra are 
displayed in Fig. \ref{Fig17}. Parameters and numerical 
results are listed in Tab. \ref{MSCu}.

\subsection{\label{SubCu2}THE CU2 LAYER}
Displacing the Cu2 atoms towards the O2,3
layer we dampen the peaks ``6.2'' and ``3.5'', but do not affect peak
``5''. Variations of the height of peak ``6.2'' by about 20\% are
found to correspond to Cu2 displacements by $\simeq 0.07$ $\rm\AA$
along the $c$\/-axis. In the range under investigation the height of
peak ``6.2'' turns out to respond almost linearly to the Cu2 positions
along $c$, $z$(Cu2). But the variations of the height of peak ``3.5'', 
which arise from the non-collinear Y--O2,3--Cu2 MS configurations 
(\#3 in Tab. \ref{Y35}), not unexpectedly, behave mildly non linearly. 
\nfig{Fig17}                             
{Calculated Fourier transform spectra, $\mid {\rm FT}(\chi k^2)\mid$,
(2-16 $\rm\AA^{-1}$) for different spacings Cu2--O2,3 ranging from
0.146-0.321 $\rm\AA$. No Gaussian broadening added. Only Cu2 is
displaced along the $c$\/-axis. The other atomic positions are fixed.
Note that the height of peak ``5'' is not affected. Note also the
shift of peak ``Cu2'' due to the varying 
Y--Cu2 bond lengths.}    
{11.968} 
{7.4375} 
{12.5}   
{t}      

\begin{table}[htb]
\begin{center}
\caption{\label{MSCu}Heights of the peaks ``3.5'' and ``6.2'' as
calculated from FEFF for different positions $z$(Cu2) along $c$.
$z$(O2,3)=0.3776, $c$\/=11.637, $a$\/=3.8091, $b$\/=3.8788 $\rm\AA$. 
Heights are in arbitrary units for $\mid {\rm FT}(\chi k^2)\mid$
from a window $k$\/=2-16 $\rm\AA^{-1}$ as used in the
analysis of the experimental data. $S_0^2$=1, 
no Gaussian broadening added.}
\begin{tabular}{ccccc} 
\hline
$\delta R$ [$\rm\AA$]&$z$(Cu2)&Dimpling [$^\circ$]
&Height ''3.5``&Height ''6.2``\\
\hline
0.146&0.365&4.4&128.4&100.5\\
0.169&0.363&5.0&139.6&107.5\\
0.204&0.360&6.1&154.2&118.3\\
0.260&0.356&7.7&171.7&134.0\\
0.320&0.350&9.5&181.9&150.1\\
\hline
\end{tabular}
\end{center}
\end{table}
In Fig. \ref{Fig17} we note that the position of the ``Cu2'' peak
shifts with the heights of peaks ``6.2'' and ``3.5''. However, a
correlation between the position of peak ``Cu2'' and the heights of
peaks ``6.2'' and ``3.5'' is experimentally not observed (see Figs.
\ref{Fig4}, \ref{Fig6}). From the experiment we find the Y--Cu2
distances are independent on the oxygen concentration ($cf.$ Tab.
\ref{YCu2Fit}, Y--Cu2 SS configurations). Therefore the Cu2
displacements along $c$ have to be constricted by fixed Y--Cu2 bond
lengths, and the Cu2 atoms may be seen to rotate with fixed radii
around the central Y sites in the [111] plane of the Y--Ba blocks.
Thereby the basal planes of the cationic sublattices (Y, Cu2, Ba) are
expanded/contracted along [110] ($cf.$ Fig. \ref{Fig18}).
Correspondingly also the $a,b$\/ lattice parameters expand/contract,
preserving or changing given local orthorhombic distortions. As
discussed in Sec. 3.1.1. we are unable to extract numbers of
the orthorhombic splittings from peak ``O2,3'', 
because the Y--O2,3 signal exhibits strong non-Gaussian disorder. 
The Y--Cu2 SS configuration is insensitive to orthorhombic 
distortions by reasons of rotational symmetry. The 
absence of non-Gaussian disorder in the Y--Cu2 SS signals
($cf.$ Fig. \ref{Fig10}) indicates that the rotational symmetry of
the Y--Cu2 SS configurations is maintained upon doping.  

According to the crystallographic data, 
see {\it e.g.} Ref. \cite{ConJil},
the $a$ axis expands upon oxyen depletion, $x_{opt}>x\rightarrow 6.5$,
whereas the $b$ axis contracts. The orthorhombicity of the unit cell
decreases correspondingly. We have included the orthorhombic
distortion and its variation with doping in 
Figs. \ref{Fig18},\ref{Fig19}.

\sfig{Fig18} 
{$Top:$\/ View from the top on the Y--\-Ba block in the unit cell of
under\-doped and op\-ti\-mum $\rm YBa_2\-Cu_3\-O_x$ (sche\-ma\-tic). Y
(small open circles) forms the top\-most, Ba (large black circles) the
lowest layer. The Cu2 and O2,3 layers are lo\-cated in bet\-ween. The
ortho\-rhom\-bic dis\-tortion is ex\-agger\-ated for rea\-sons of
clarity. The in\-crease of the ortho\-rhom\-bicity on doping oc\-curs
by con\-trac\-tion of the $a$\/-, and ex\-pansion of the $b$\/-axis.
Ar\-rowed lines con\-nect the atoms of the Y--Cu2--Ba three body
scatte\-ring con\-fi\-gu\-rations (Peak ``6.2''). 
$Bottom:$\/ Side view of the doped
Y--\-Ba blocks along [110] with con\-stant Y--\-Cu2 bond lengths. On
doping Cu2 moves along $c$ to\-wards the Ba layer there\-by
de\-creasing the for\-ward scat\-tering angle from 6$\rm^{\circ}$ to
4$\rm^{\circ}$. The weak ex\-pansion along c with do\-ping is
ex\-pec\-ted from the crys\-tallo\-gra\-phic struc\-ture.} 
{12.5}     
{7.718}   
{13.53}   
{3.8}      
{t}        

The congruent $T$\/-dependencies ($T>80$ K) of peak ``5'' for all
concentrations ($cf.$ Fig. \ref{Fig13}) lead to the conclusions, that
not only the distribution of Y--O2,3\---Ba forward scattering angles
(\#2--\#6 in Tab. \ref{Y5}) remain unaffected upon doping, but also
the Y--Ba vibrational frequency.

\sfig{Fig19} 
{$Top:$\/ View from the top on the Y--\-Ba block in the unit cell of
under\-doped and op\-timum $\rm YBa_2\-Cu_3O_x$ (sche\-ma\-tic). Same
cell as in Fig. \ref{Fig18}, but rotated by $\simeq 45\rm^{\circ}$ 
Ar\-rowed lines con\-nect the atoms of the Y--O2,O3\---Ba three body
scatte\-ring con\-fi\-gu\-rations (Peak ``5''). 
$Bottom:$\/ Side view of the doped Y--\-Ba blocks along [010] with 
an Y--O2,3--\-Cu2 for\-ward sca\-tter\-ing angle 
$\simeq 13\rm^{\circ}$ kept fixed on do\-ping.   
Upon do\-ping the po\-sitions of the pla\-nar oxy\-gens 
do not move (or only very weak\-ly) along $c$ to\-wards the Ba
layer, but per\-pen\-di\-cu\-lar to $b,a$. 
There\-by the Y-O2,3 bond lengths ex\-pand/\-con\-tract with\-in 
the margins of the ortho\-rhom\-bic
dis\-tor\-tion. Dashed hori\-zon\-tal lines indi\-cate the 
po\-sitions of Cu2 from Fig.
\ref{Fig18}. Same ex\-pan\-sion along 
$c$ as in Fig. \ref{Fig18}.} 
{12.5}     
{6.927}   
{11.169}   
{4.6}      
{t}        

Therefore we may safely attribute the variations of peak ``6.2''
observed on doping to different forward scattering geometries in the
Y--Cu2\---Ba MS configurations, and we may discard effects from
the doping dependence of the Y--Ba vibrational dynamics. Moreover, if
we take into account the static Ba displacements from crystallographic
work, we find peak ``6.2'' of the underdoped sample even more strongly
dampened than in the case of unaltered Ba positions.

Also the significant difference between the temperature behaviours of
peak ``6.2'' in the ``underdoped'' sample on the one hand side, and
the optimum/overdoped samples on the other hand side ($cf.$ Fig.
\ref{Fig11}), may be attributed to different Cu2 positions along $c$.
The staircase-like $T$\/-dependence of peak ``6.2'' in the
``underdoped'' sample indicates from the Cu2 displacements another
characteristic temperature than $T_c$. Around $T^*=160$ K (close to
the so-called spin gap temperature) the Cu2 atoms approach more
closely to their positions in the optimum/overdoped samples.

\subsection{\label{subsubO23}THE O2,3 LAYER}
The height of peak ``5'' is expected to depend 
sensitively on the forward scattering
angle in the Y--O2,3--Ba MS configurations, but weaker than in the 
Y--Cu2--Ba MS (peak ``6.2'') configurations. However, due to the 
orthorhombic unit cell the height of peak ``5'' measures 
a distribution of Y--O2,3--Ba forward scattering angles. 
It turns out that the distribution of 
Y--O2,3\---Ba forward scattering angles
is almost unaffected upon doping, if we
allow the Y--O2,3 distances to expand/contract within the 
large error margins of $\simeq 1$\% ($cf.$ Tab. \ref{O23Fit}) 
set by the SS analysis of peak ``O2,3''. 
The resulting displacement of $e.g$ O2 along $c$ is negligibly 
weak and indicated in Fig. \ref{Fig19} ($bottom$) by the drawn out 
horizontal lines labelled $\delta\bar{z}\simeq 0$. The other way
round, if we fix the average Y--O2,3 distance, the average forward
scattering angle varies upon doping by $\simeq 3\rm^{\circ}$, which
clearly pushes the height of peak ``5'' beyond the error limits of the
experimental data. From the congruent data sets in Fig. \ref{Fig13} we
conclude that the O2,3 layer does not move along $c$ for 
$6.801<x\leq x_{opt}$. But the O2,3 atoms have large degrees of 
freedom perpendicular to the Y--O2,3--Ba scattering geometry, $i.e.$ 
along $a$ and $b$, respectively (Fig. \ref{Fig19}). 

It should be noted that a constant distribution of 
Y--O2,3\--Ba forward scattering angles, given as an experimentally
given constraint, might be also achieved 
by rotating the Y--O2,3--Ba scattering triangle around the base 
Y--Ba path. Such rotations, however, would lift, at least partially, 
the degeneracies of the forward scattering configurations and alter 
their angular distributions. It is interesting to note that 
the shapes of peaks ``O2,3'' and ``5'' in the calculated  
Fouriertransform spectra (Fig. \ref{Fig17}) show
appreciable differences in comparison with the experimental ones
(Figs. \ref{Fig6},\ref{Fig4}). Apparently the geometry of the 
Y--O2,3 cluster is only poorely described by the 
crystallographic data used for the calculations and
listed in Tabs. \ref{YO23},\ref{Y5}.

\subsection{\label{subMax}THE CU2--O2,3 LAYERS}
\nfig{fig20}                                        
{The underdoped--overdoped transition 
in $\rm YBa_2Cu_3O_x$ at $T=25\pm 5$ K as determined from 
the Y--Cu2--Ba (``6.2'') and Y--O2,3--Ba (``5'') 
MS scattering. $Left:$\/ The heights of the peaks ``6.2'' 
(closed circles) and ``5'' (open circles) as a function of oxygen
concentration, $x$. The data points from the two 
underdoped samples, $x=6.826$ and 6.886, are included from unpublished
results \cite{Unpub}. Lines connecting the data points are guides to
the eyes. $Right:$\/ Ratio of the peak heights shown left.}    
{12.096} 
{6.084}  
{12.5}   
{t}      
\noindent
In Fig. \ref{fig20} {\it (Left)}\/ we plot the 
heights of peaks ``6.2'' and
``5'' at $T=25\pm 5$ K $vs.$ the oxygen concentration $x$. 
Peak ``6.2'' increases strongly with $x$ indicating 
that the Cu2 layer shifts along $c$ towards the Ba layer. 
Up to $x\simeq x_{opt}$ the height of peak ``5'' is almost
independent on $x$, but it increases by 
$\simeq 10$ \% at $x=6.984$. From
the preceding discussion it is tempting to relate the difference of
the two peak heights directly to the spacing of the Cu2--O2,3 layers.
However, the configurational average of the Y--O2,3--Ba MS scattering
signals and thus the height of peak ``5'' were shown to depend
sensitively on the orthorhombicity and the anharmonic O2,3 dynamics,
which also depend on $x$. If we assume negligible changes of 
the anharmonic O2,3 dynamics in the underdoped regime, 
we may read from Fig. \ref{fig20} {(\it Left)}\/ 
the average Cu2--O2,3 spacing (or dimpling of the $\rm
CuO_2$ planes) to increase for $x\rightarrow x_{opt}$.

A recent Raman study of the oxygen
vibrations across the underdoped-overdoped transition \cite{LiaMue} 
reports for the overdoped regime, $x>6.93$, a relative softening 
of the O2,3 in-phase vibrations in $z$\/-direction (440 $\rm cm^{-1}$, 
$\rm A_g$) by $\simeq 40$\%. Taking into account such strong 
changes of the anharmonic vibrational O2,3 dynamics 
we expect the relationship between the height of peak ``5'' and
average O2,3 position along $c$ to be quantitatively but not
qualitatively different from that in the underdoped regime. Therefore
we read from the data at $x=6.984$ that the heavily overdoped O2,3
layer moves along $c$ in the same direction as the Cu2 layer and
thus reduces the Cu2--O2,3 spacing. The ratio of the peak heights
depicted in Fig. \ref{fig20} {\it (Right)}\/ exhibits a clear maximum
at optimum doping. It indicates a structural phase transformation of
the displacive type at this notable point in the phase diagram. Doping
in the overdoped regime displaces the average position of the O2,3
layer along $c$ towards the Cu2 layer and thereby reduces the Cu2--O2,3
spacing. In the underdoped regime the Cu2--O2,3 spacing is reduced by
oxygen depletion, but inversely by displacements of the 
Cu2 layer along $c$ towards the O2,3 layer.

\section{\label{SecSumm}Concluding Remarks and Summary} 

We have extracted local structural information from systematic
measurements of the Y-$K$ EXAFS of $\rm YBa_2Cu_3O_x$ for
$x=6.801-6.984$ and at $T=20-300$ K. The structural analysis was
focussed on the Y--O2,3 and Y--Cu2 single scattering configurations,
and the multiple scattering configurations Y--O2,3--Ba (peak ``5'')
and Y--Cu2--Ba (peak ``6.2''). With respect to the large amount of
structural anomalies, directly visible in the Fouriertransform spectra
as a function of doping and of temperature, the structural analysis
presented in this lecture is far from being complete, and in part 
still preliminary. In particular the important multiple scattering 
signal demand for a quantitative analysis including the 
vibrational dynamics. To summarize, we emphazise the
following results:

\begin{enumerate}

\item The Y--Cu2 bondlengths are independent on doping. In the normal
phase the Y--Cu2 pairs vibrate harmonically, but in the
superconducting phase the Y--Cu2 vibrations freeze out. On doping the
Cu2 atoms shift along the $c$\/ direction towards the Ba layer. 

\item The Y--O2,3 pairs exhibit strong anharmonicities. The degree of
non-Gaussian disorder and the strong anharmonic vibrational dynamics
depend significantly on the oxygen concentration. The average Y--O2,O3
bondlengths are almost independent on doping (within the error limits
set by our harmonic analysis). The Y--O2,3 mean-squared deviations do
not freeze out in the superconducting phase as the Y--Cu2 vibrations.
But the mean-cubic deviations of the Y--O2,3 pairs are large, depend
on doping, and exhibit a clear singularity at $T_c$ (shown for
x=6.968).

\item In the underdoped regime the average spacing O2,3--Cu2 
increases with oxygen concentration, $x\rightarrow
x_{opt}$, due to relative displacements of Cu2 towards Ba. 
The position of the O2,3 along $c$ is almost unaffected by doping. 
Staircase-like displacements of Cu2 along $c$ occur at characteristic
temperatures other than $T_c$. The average position of the 
O2,3 layer along $c$ behaves monotonously as a function of 
temperature.

\item In the overdoped regime increasing oxygen concentration 
displaces the Cu2 layer along $c$ further towards Ba. 
But the O2,3 layer exhibits a clear tendency to shift in the same 
direction along $c$ thus decreasing the 
average O2,3-Cu2 spacing. 
No staircase-like Cu2 displacements are observed in the normal phase.  

\item Optimum doping is a notable point in the phase digram of $\rm
YBa_2Cu_3O_x$, also concerning the structural degrees of freedom. 
Here the O2,3--Cu2 spacing is largest, and the relative displacements
between O2,3 and Cu2 invert upon doping. In the optimum doped 
superconducting phase the Cu2 position along $c$ is independent on 
temperature for $T<80$ K.

\end{enumerate}

\noindent

\begin{acknowledgements}
Part of this work has been performed during a stay of 
J. R\"ohler as a visiting scientist at the ESRF. He is grateful
to B. Lengeler and to the staff of the ESRF for support. 
Beam time was under the proposal HC362. The help of J. Jensen
and A. Filipponi during the data collection is gratefully
acknowledged.
\end{acknowledgements}

\end{document}